# Search for eV neutrino sterile: Status of STEREO experiment

**Ilham El Atmani** [1,†] **, On behalf on the STEREO Collaboration**

[1] IRFU, CEA, Université Paris-Saclay, 91191 Gif-sur-Yvette, France.
† Present address: Hassan II University, Faculty of Sciences Ain Chock, BP 5300 Maarif, Casablanca 20100, Morocco.

Email: ilhamelatmani314@gmail.com

**Abstract.** Reactor neutrinos have played a key role in understanding neutrino physics since their discovery. The so-called reactor-anti-neutrino-anomaly RAA, a ~6.5% deficit of the mean observed neutrino flux compared to the prediction appeared recently. This anomaly could be interpreted by the existence of a fourth, sterile, neutrino and this hypothesis is currently being tested by the very short baseline experiment STEREO. The latter is installed at very short distance (9 -11m) from the compact core of the ILL research reactor in Grenoble-France and collecting data since November 2016. The ILL core is highly enriched in $^{235}$U and releases a nominal thermal power of 58.3 MW. The geometry of the STEREO detector, segmented into six identical cells filled with Gd-loaded liquid scintillator, is designed for a direct test of a new oscillation pattern in the L/E range around 1 m/MeV, relevant for the RAA. First published results of STEREO have demonstrated the mitigation of the background induced by the reactor and the cosmic-rays and a good energy response. The data taking is now in progress with very stable conditions favorable for an improved accuracy. We will present an overview of the experiment and an update of the sterile neutrino analysis. A refined prediction of the neutrino spectrum emitted by the ILL reactor is also presented.

## 1. Introduction

A significant deficit (6.5) % of the anti-neutrino flux between all measurements and prediction, known as reactor anti-neutrino anomaly (RAA), can be interpreted as a new short distance oscillation toward a sterile neutrino with a mass around 1 eV. This is STEREO's first motivation to check this hypothesis independently from any reactor prediction, by comparing the neutrino spectra detected in 6 identical cells. However the RAA could also be explained as an error on the flux prediction. Indeed the reactor experiments (Daya Bay [1], Reno [2] and Double Chooz [3] ) recently observed a similar deviation (spectral shape) from the previous prediction in the 4-6 MeV energy region. In addition Daya Bay results suggest that the RAA is mainly driven by $^{235}$U, with a rate deficit of 8% for this isotope while $^{239}$Pu would be on expectation. STEREO will measure the pure $^{235}$U neutrino fission spectrum, as a second motivation, in order to identify the primary contributor to the RAA.

## 2. STEREO experiment

The STEREO experiment [4] is located at 9-11m from ILL research reactor core in Grenoble (France), which produces a large ($10^{19}$ s$^{-1}$) pure flux of electronic anti-neutrinos $\bar{\nu}_e$, with a nominal thermal power of 58.3 MW and about 3 cycles per year each of 50 days. The high enrichment of 93

% $^{235}$U fuel allows to provide a new reference anti-neutrino spectrum for pure $^{235}$U fission, knowing that we have no damping of the amplitude of the oscillation thanks to the compactness of the reactor core ($\Phi \approx 40$cm). However the mitigation of background, mainly coming from neighbouring experiments using neutron beams around us is challenging. Fortunately, the location of the STEREO detector under a water channel (15m.w.e) helps reducing the flux of cosmic rays, second important source of background. Thanks to the large passive shielding added all around the detector and the rejection of external background provided by the muon veto on top of the detector and the Gamma-Catcher crown of liquid scintillator around the target, a Signal-to-Noise ratio of about 1 is achieved. The target volume is filled with Gd-loaded liquid scintillator (LS-Gd) allowing a good energy resolution and efficient neutron capture.

## 3. Selection event

As most of the reactor neutrino oscillation experiments, we detect the anti-neutrino in (LS) through inverse beta β decay (IBD) process $\bar{\nu}\,p \rightarrow n\,e^+$ with a threshold at 1.8 MeV. A neutrino candidate is selected when a prompt (positron) and a delayed (neutron capture) events are detected with energy in [1.6, 7] MeV and [4.5, 10] MeV respectively. The delayed energy corresponds to the neutron capture on Gd. Both events are separated by a correlation time between 2μs < Δt < 70μs adapted to the ~16μs neutron capture time. The muon induced background is further rejected by imposing isolation cuts at 100 μs before and after the pair candidates. Additional topological cuts require that less than 1 MeV is deposited in the neighbouring cells of the prompt event and less than 0.4 MeV in the other cells.

## 4. Energy calibration

The calibration is important for an accurate energy scale, knowing that any uncertainty in the latter will reduce sensitivity. The determination of the reconstructed energy $E_{rec}$ has been obtained from the PMT collected charges Q in each cell using an inverse matrix formalism to correct for light leaks between cells. It is based on regular calibration with radioactive γ and neutron sources deployed inside six cells at five heights along z-axis, The fine tuning of the calibration and light leaks coefficients is performed with a $^{54}$Mn source ($E_\gamma$ = 0.835 MeV). Figure 1 illustrates the very good agreement achieved between Data and Monte-Carlo for this reference energy distribution. The only sizeable discrepancy occurs below 300 keV due to trigger threshold effects not fully implemented in the MC (not necessary because this threshold is very far from the software threshold of 1.6 MeV applied in the analysis). This anchoring of the detector response is performed every week, allowing an accurate monitoring of the detector response. Away from the $^{54}$Mn energy point the well-known quenching effect, inherent to liquid scintillator, induces some non-linearity in the detector response. It is accurately tuned in the MC using $k_B$ parameter of the Birk's model [5]. Then to evaluate the volume effects we use an independent cross-check from the reconstructed neutron capture on H ($E_{n-H}$ = 2.22 MeV) coming from the simulation of cosmic events, the expected vertex distribution of these events is quite uniform in the detector therefore probing the detector response in a way that is complementary to the calibration data and closer to the expected distribution of neutrino events. Figure 2 shows the evolution in time of the position of the reconstructed n-H peak, fitted with a crystal ball function. Sub-percent stability is achieved.

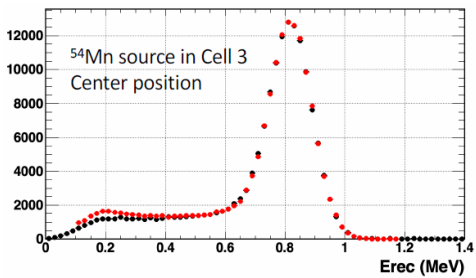 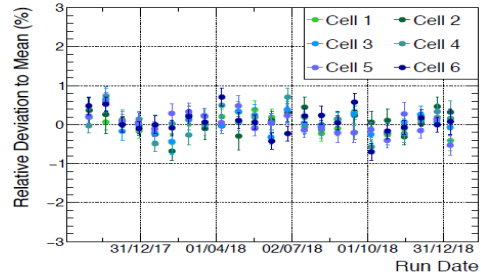

**Figure 1.** Comparison between Data (black points) and simulation (red points).

**Figure 2.** Time stability of n-H peak position

## 5. Detection efficiency and Improvement of the description of the Gd cascade

The selection cuts discussed in section 3 lead to a global detection efficiency of 61.4 ± 0.9%. The main source of uncertainty in this number comes from the neutron detection efficiency. It is carefully studied with an Am-Be source deployed in all cells. The main challenge of this analysis is the description of the complex cascade of gamma rays emitted after the neutron capture on Gd. In a small detector like STEREO the amount of energy leakage out of the liquid scintillator becomes very sensitive to the energy and multiplicity distributions in the cascades. A significant improvement of the agreement between data and MC was achieved by replacing the cascades generated by the standard GEANT4 libraries with the predictions of the FIFRELIN code [6], solving a long standing issue observed in several neutrino experiments. Ten millions of cascades have been made available to the community on a zenodo repository [7].

## 6. Background stability and Neutrino extraction

The accidental background is efficiently reduced by the shielding. The residual contribution is accurately measured online via off-time coincidence windows and thus subtracted without significant systematic uncertainty. The correlated background, induced by the cosmic rays, remains the main component. The Pulse Shape Discrimination (PSD) capability of the liquid scintillator is a crucial asset to reject further this background. As illustrated in Figure 3, taking the ratio of the late charge $Q_{tail}$ to the total charge $Q_{tot}$ of the pulse shows two populations of events, the neutron induced proton recoils at high PSD and the "electron like" events (including the neutrinos) at smaller PSD. The red points illustrate the PSD distribution when the reactor is OFF. This background model is used to fit the reactor ON data with a global "$a$" norm factor to account P$atm$ effects, plus one Gaussian for the neutrino candidates, which are extracted per cell and 500 keV energy bin (green), including the distributions of accidentals pairs (grey) with high statistics . It was shown that the shape of the reactor OFF model was very stable, even when splitting on purpose the data into extreme bins of atmospheric pressure or water level in the reactor pool. This combined with extensive measurements of the background spectrum (more reactor OFF data have been acquired than reactor ON data), insures an accurate extraction of the neutrino signal.

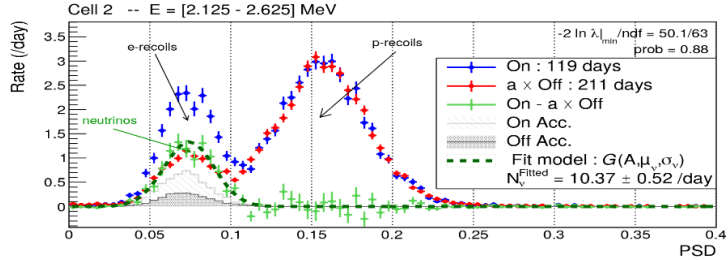

**Figure 3.** Neutrino Spectrum extraction from PSD

## 7. Oscillation analysis

The extracted neutrino rates as described in section 6, give the six spectra in figure 4, corresponding to the number of neutrinos $D_{l,i}$ measured for all cells $l$ and energy bins i with their statistical uncertainty $\sigma_{l,i}$. They are compared to a common spectrum shape defined as the expected numbers of neutrinos $\phi_i * M_{l,i}$ representing an oscillation model $M_{l,i}(\mu, \sigma, \vec{\alpha})$ with $\phi_i$ set as eleven free normalization parameters, one for each energy bin i, but common the all six cells and insensitive to the prediction. All systematics of normalization and energy scale are described by $\vec{\mu}$ and $\vec{\alpha}$ set of nuisance parameters also known as pull terms. The tested $\chi^2$ is written as:

$$\chi^2 = \sum_{1}^{N_{Cells}} \sum_{i}^{N_{Ebins}} \left( \frac{D_{l,i} - \Phi_i M_{l,i}(\mu, \sigma, \vec{\alpha})}{\sigma_{l,i}} \right)^2 + \sum_{1}^{N_{Cells}} \left( \frac{\alpha_l^{Norm\ U}}{\sigma_l^{Norm\ U}} \right)^2 + \left( \frac{\alpha^{Escale\ C}}{\sigma^{Escale\ C}} \right)^2 + \sum_{1}^{N_{Cells}} \left( \frac{\alpha_l^{Escale\ U}}{\sigma_l^{Escale\ U}} \right)^2 \qquad (1)$$

As a cross-check an independent analysis is also conducted with a $\chi^2$ formula based on the covariance matrix formalism for the systematics. In both cases the "true" PDF's of the $\Delta\chi^2$ are built by generating thousands of pseudo-experiments for each bin of the (Δm², sin²(2θ)) plane. The "no-oscillation" hypothesis being not rejected by the data, a rejection contour is determined using the raster scan method. It illustrated in Figure 5 for the phase I+II data. The curve is centered around the expected sensitivity contour, with oscillations induced by the statistical fluctuations of the data points. The initial best-fit of the RAA is rejected at more than 99% level.

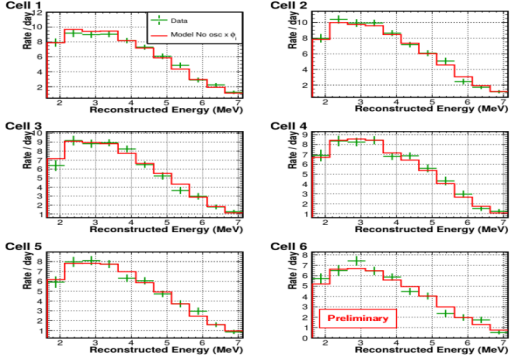
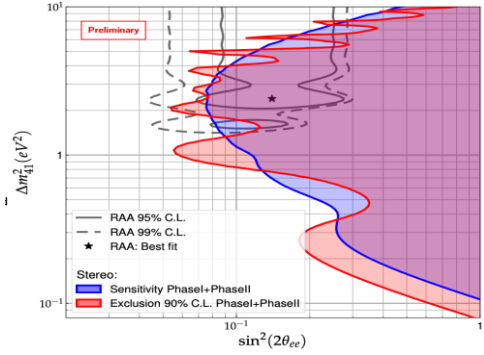

**Figure 4.** Comparison of measured (green) and simulated (red) spectra.

**Figure 5.** Exclusion contour in the oscillation parameter (red) and sensitivity (blue).

**8. Prediction anti-neutrino spectrum**

Nuclear reactor ILL produces a large ($10^{19}$ s$^{-1}$) $\bar{\nu}_e$ flux by $\beta$-decay chains of the fission. Reactor parameters are controlled and used as essential ingredients, for instance the reactor thermal power $P_{th}$ that presents the most dominant uncertainty ($\partial P_{th}/P_{th} \pm 1.4\%$) to calculate the total $\bar{\nu}_e$ flux emitted:

$$S(E,t) = \frac{p_{th}(t)}{\sum_k f_k(t) E_k} \sum_k f_k(t) S_k(E)_{corr} \quad (2)$$

Where $f_k(t)$ the fission fraction of the isotope $k = \{^{235}U, ^{239}Pu\}$, $S_k$ the $\bar{\nu}_e$ spectrum from prediction model equation (3) $\pm 2.4\%$ [8], $<E_k>$ the mean energy release per fission evaluated in ILL case.

*8.1. Method*

The Huber model [8] is a standard reference of the anti-neutrino spectrum $S_k$ after 12 h irradiation time of pure $^{235}U$, obtained from the conversion of the ILL total β-spectrum, available in 50 keV. In order to provide an accurate prediction, we take into account several corrections beyond the dominant contribution of the pure fission spectrum predicted for pure $^{235}U$. From the FISPACT reactor simulation a mean fission fraction of 0.7% comes from the $^{239}Pu$ isotope during a typical ILL cycle. This reduces the predicted neutrino rate by about 0.3%. In the following we will discuss several corrections affecting the low energy part of the predicted spectrum: Residual neutrinos from spent fuel $C_{spent\,fuel}$, off equilibrium effects $C_{off\text{-}equi}$ and neutrino from $^{28}Al$ neutron capture $C_{Al}$. The emitted anti-neutrino spectrum is then given:

$$S_k(E)_{corr} = \sum_k f_k S_k(E).(1 + C_{spentfuel} + C_{offequi} + C_{Al}) \quad (3)$$

*8.2. Neutrino from Spent fuel*

At the end of a reactor cycle the control rod is fully inserted in the core and the fission reactions stop. However the long-lived fission fragments accumulated in the core keep emitting neutrinos via their beta-decay chain. A careful estimation of this residual flux has been performed, requiring the simulation of the inventory of the fission products along the cycle as well as the monitoring of the position of the spent fuel elements after the end of their cycle.

After the reactor stop the fuel element cools down for 1 day at its nominal position, then it is moved (under water) in a dedicated chimney in the pool. After 50 days the residual power a small enough to move the spent fuel to a storage area, which turns out to be in the water channel, on top of the STEREO detector. Figure 7 illustrates the history of the predicted neutrino flux detected in STEREO and due to a single spent fuel element, the start time being the reactor stop. This flux is expressed in percentage of the detected flux when the core is operated at nominal power (55 MW). As expected the residual flux is dropping very fast below the 1% level before reaching the small contribution of fission products with year scale half-life. However the above mentioned changes of the location of the spent fuel each time bring it closer to the STEREO detector, amplifying the detected flux. This is clearly visible in figure 6 after 1 day and 50 days. Then this model is used to compute the cumulative flux from all spent fuel elements taking into account the dates and mean power of all previous cycles (Figure 7). In order to mitigate this residual flux the 24h following a reactor stop are always removed from the reactor OFF data taking. Finally, the flux of residual neutrinos was found small and further suppressed by using subtraction $C_{Spentfuel}^{U5} = \int_{Start}^{Stop+1} ON - \int_{Start}^{Stop+1} OFF$ , thus the correction amount less than (~ 0.2 ) % in the first energy bin and roughly the half in the second (Figure 10), the uncertainty was neglected due to the tiny contribution.

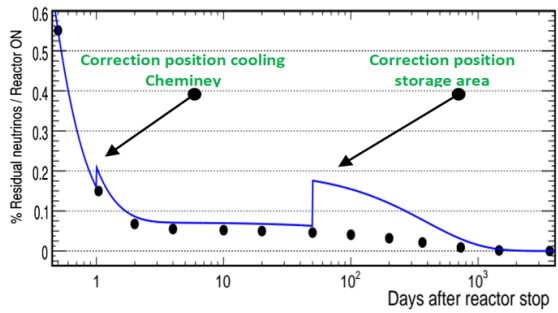

**Figure 6.** One spent fuel element at different position % residual neutrinos to Reactor ON.

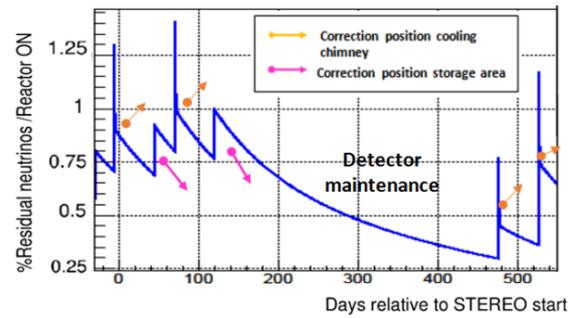

**Figure 7.** Monitoring of all elements of % residual neutrinos to Reactor ON ( STEREO).

*8.3. Off equilibrium effects*

Similarly to the residual $\bar{\nu}$ flux after a reactor stop, the long lived fission products will take some time to reach equilibrium (production rate = decay rate) after the start of a cycle. Given the fact that the reference Huber spectrum is based on measurements taken after only 12h to 1 day of irradiation of a U or Pu foil, off-equilibrium corrections have to be taken into account. In practice the long-lived isotope will keep accumulating after this irradiation time and an increase of the spectrum is expected due to this effect. It concerns only the low energy part of the spectrum, since there is a strong anti-correlation between the endpoint energy of the beta-transitions and their lifetime. To compute this correction the FISPACT evolution code, predicting the isotope inventory versus time, was coupled to the BESTIOLE code, providing the beta-spectra of all fission products. The correction is found to be at the 1% level in the first bin of the STEREO analysis, equivalent to [2.4, 2.9] MeV in neutrino energy (see figure 8). Then it decreases rapidly with energy. The amplitude of this correction is smaller than previous calculations applied to the case of commercial reactors because the ILL cycles are 50 days only, preventing large accumulation effects. A safe 30% relative uncertainty is associated to this correction.

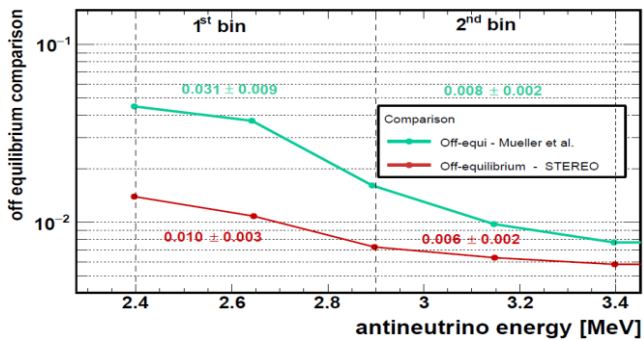

**Figure 8**, Comparison between STEREO and Mueller *et al.* [9] Off equilibrium in 250 keV.

*8.4. Neutrinos from Neutron capture ($^{28}$Al)*

Last but not least, a contribution to the neutrino flux can come from the capture of neutrons on other elements than uranium, generating neutron rich nuclei that can undergo beta-decay. In the case of the ILL core the dominant element by far is $^{27}$Al, present in the uranium oxide itself, in the mechanics of the core, in the neutron guide and the heavy water vessel around the core. The formed $^{28}$Al isotopes undergo beta decay with a 2.2 min half-life and 2.863 MeV neutrino endpoint (above STEREO analysis threshold). A refined TRIPOLI simulation of the ILL reactor (figure 9), including all the main mechanical elements has been performed showing that on average 0.343 neutron per fission is captured on $^{27}$Al, corresponding to 11% of the total neutron flux. This quite large contribution induces the main correction to the first energy bin of STEREO: 10.3% increase of the predicted flux (figure 10). The second bin is already above the endpoint hence no other correction apply. The neutron balance being a very sensitive parameter of the TRIPOLI simulation the relative uncertainty on the $^{27}$Al is small, bellow the 1% percent level from the simulation itself. The uncertainty on the total mass of Al defined in the TRIPOLI geometry should be added in quadrature, its estimation is currently in progress.

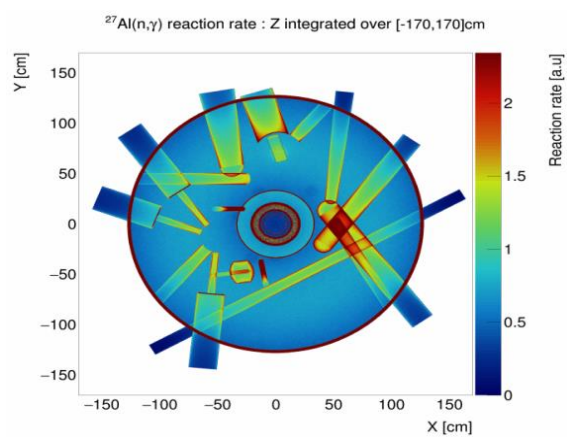
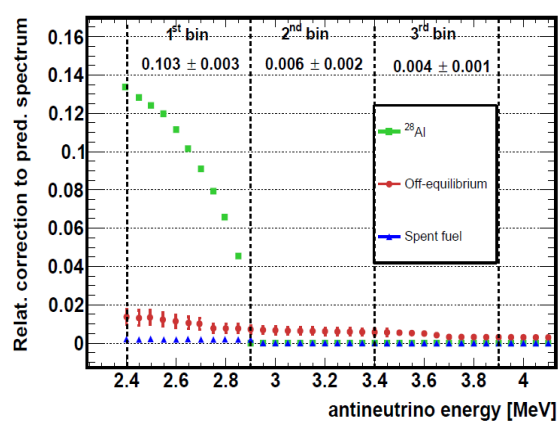

**Figure 9.** Top view of $^{27}$Al ($\gamma$,n) rates over reactor core vessel.

**Figure 10.** Relative corrections to Huber model spectrum.

**Acknowledgments**

This work is funded by the French National Research Agency (ANR) within the project ANR13-BS05-0007 and the "Investments for the future" programs P2IO LabEx (ANR-10-LABX0038) and ENIGMASS LabEx (ANR-11-LABX-0012). We are grateful for the technical and administrative support of the ILL for the installation and operation of the STEREO detector. We further acknowledge the support of the CEA, the CNRS/IN2P3 and the Max Planck Society..